\pdfoutput=1
\documentclass[conference]{IEEEtran}
\usepackage[utf8]{inputenc}
\usepackage[T1]{fontenc} 
\usepackage{amsmath}
\usepackage{booktabs}
\usepackage{url}
\usepackage{pgfplots}
\pgfplotsset{compat=1.17}
\providecommand{\doi}[1]{doi:#1} 

\begin{document}

\title{No Edges, No Verdict: A Large-Scale Empirical Study of Declared
       Dependency Graphs in 78K SBOMs in the Wild}

\author{\IEEEauthorblockN{Artur Zi\k{e}ba-Kozarzewski}
\IEEEauthorblockA{0000-0003-4696-4983\\
KRYPTON Polska Sp.~z~o.o.\\
Warszawa, Poland\\
Email: a.zieba@krypton-polska.com}}

\maketitle

\begin{abstract}
Software Bills of Materials (SBOMs) are read not only as inventories but as
dependency graphs: vulnerability triage, reachability filtering and impact
analysis traverse the edges an SBOM declares. We present the first
population-scale study of the declared graph, over 78,612 SBOMs committed to
public code (77,092 parseable). Of these, 52.9\% omit the dependency-relationship
field, although the NTIA minimum elements require it (42.3\% once the 46.5\% of
files carrying the tool entry of a single generation harness are set aside), and
a further 8.8\% declare a graph in which most components are orphans, so for
61.7\% a missing path says nothing about reachability. Scans of 11 large
container images are all degenerate (91.4--98.0\% orphans), through file entries.
The format does not predict which documents these are; the producing tool does
($\eta^2$ = 0.41--0.65), jointly with the ecosystem where a document describes
one. CycloneDX completeness declarations appear in 0.10\% of CycloneDX
documents, and a widely used quality score separates the regimes only weakly.
Reading such graphs closed-world, so that no declared path means unreachable,
discards almost the whole inventory: under a production detector the median
degenerate document would lose 96.4\% of its components, which carry 29
version-confirmed known-exploited CVEs across 1,901 documents. Where graphs are
informative, the veto touches 14 such CVEs across 164 documents, which measures
its reach, not its correctness. On an 18-SBOM testbed scored with the KEV flag
removed (EPSS retained), the veto suppressed 5 of 14 version-confirmed
known-exploited instances, Log4Shell among them; reading silence as
\emph{unknown} restores them in this illustration. We release the scanner and the
dataset.
\end{abstract}

\section{Introduction}\label{sec:intro}

SBOMs promise transparency for the software supply chain, and a growing tool
ecosystem consumes them for vulnerability management. Several of the
highest-value consumption scenarios read the SBOM as a \emph{graph}: which
components does the application actually pull in, through which paths, and is a
vulnerable component reachable from the artifact's roots? The NTIA minimum
elements make this explicit: \emph{dependency relationships} are a required
field, not an optional enrichment~\cite{ntia2021minimum}.

This paper asks, at population scale, \textbf{what the graph that SBOMs actually
declare looks like}. Prior large-scale studies audited SBOM \emph{fields}
(standards adherence, licence accuracy, inter-tool
consistency~\cite{wang2026adherence}) or the gap between the files in an
artifact and its metadata~\cite{reddypalle2026sdm}, that is, what the SBOM does
not declare at all. The one study that measured the declared graph did so on 119
documents~\cite{nocera2025adoption}. We study a different axis: \emph{what the
SBOM declares, but without edges}.

The consumer inference at stake, which we call the \emph{veto}, works on two
levels. As a reading of the document, it is the step from ``the declared graph
has no path from the root to component $c$'' to ``$c$ is unreachable'', and from
there to setting $c$'s vulnerabilities aside. This treats the declared graph as
complete, a \emph{closed-world} reading~\cite{reiter1978closed}. Inside a
prioritization system the same step becomes a factor: the system we study
multiplies each vulnerability's score by a reachability factor of 1 for a
confirmed component, 0.5 for a possibly reachable or unknown one, and 0 for an
unreachable one. The \emph{open-world} alternative maps a missing path to
\emph{unknown}, and a degeneracy detector decides when a graph is too sparse for
a missing path to count as evidence.

Scanning all 78,612 SBOMs of the Wild SBOMs corpus~\cite{soeiro2025wildsboms},
we find the declared graph missing far more often than a consumer would expect.
Of the parseable SBOMs, 52.9\% omit the dependency-relationship field, and a
further 8.8\% declare a graph in which most components are orphans. For 61.7\% of
the documents, then, a missing path says nothing about reachability. One
generation harness accounts for at least 36,551 documents; with it set aside,
42.3\% lack a dependency block (Section~\ref{sec:rq1}). The closed-world reading
is unsound because SBOM extractors are
\emph{soundy}~\cite{livshits2015soundiness} in one direction
only: we take a declared edge to be almost always a true dependency (an
assumption we do not measure), but a missing edge is not evidence of
independence. The specification even provides an escape hatch, CycloneDX
\texttt{compositions}~\cite{cyclonedx2024spec}, yet only 59 of 61,691 CycloneDX
SBOMs in the wild (0.10\%) use it.

We structure the study around six questions:
\begin{itemize}
  \item \textbf{RQ1} What dependency graph do published SBOMs actually declare,
    and how does the population divide into regimes (taxonomy; presence of the
    NTIA dependency-relationship field)?
  \item \textbf{RQ2} Which of format, generating tool, or ecosystem of the
    described software \emph{predicts} the regime, and how much variance does each
    account for?
  \item \textbf{RQ3} Can a consumer read the graph's absence off
    specification-level completeness declarations or field-quality scores?
  \item \textbf{RQ4} What would a consumer who reads these graphs closed-world
    remove across the population? \textbf{RQ4.1}: where the graph is
    uninformative. \textbf{RQ4.2}: where it is informative.
  \item \textbf{RQ5} What does the closed-world reading cost in a
    vulnerability-prioritization system?
  \item \textbf{RQ6} Does an open-world reading restore what the veto removes?
\end{itemize}
RQ1--RQ4 are answered on the wild corpus (Sections~\ref{sec:population}
and~\ref{sec:trap-population}); RQ5 and RQ6 on an 18-SBOM testbed, as an
illustration of the mechanism (Section~\ref{sec:trap}). In that testbed, scored
with the KEV flag removed, the veto suppressed 5 of the 14 known-exploited
(KEV-listed) instances whose package version we could confirm, Log4Shell among
them.

\textbf{Contributions.}
\begin{itemize}
  \item \textbf{C1 --- population characterization.} The first topology census of
    declared dependency graphs at population scale (78,612 SBOMs), with a
    three-way taxonomy (\textsc{no\_edges} / \emph{degenerate-with-edges} /
    \emph{connected}), the NTIA dependency-relationship field measured
    population-wide (absent from 52.9\%), a variance decomposition in which the
    generating tool, not the format, predicts edge emission (jointly with the
    ecosystem where a document describes one), and the measured use of
    completeness declarations (0.10\%) and quality scores.
  \item \textbf{C2 --- semantics.} The asymmetric soundness argument, an
    \emph{unknown} reachability level distinct from evidence-based
    \emph{unreachable}, and a degeneracy detector (in-degree orphan ratio $>0.5$)
    deciding when the veto is invalid.
  \item \textbf{C3 --- what the veto removes.} Over the corpus, in documents
    where the graph is uninformative, the veto would remove components carrying
    29 KEV-listed CVEs confirmed at the exact version across 1,901 documents; the
    mirror measurement where the graph is informative gives 14 CVEs across 164
    documents, reported as an extent, not a verdict on correctness. In our own
    prioritization system the veto suppressed 5 of 14 version-confirmed KEV
    instances, presented as an illustration of the mechanism, not a measurement
    of the repair.
  \item \textbf{C4 --- artifact.} A streaming scanner (no decompression of the
    5\,GB corpus), the full per-SBOM topology dataset, the 11-image corpus and the
    frozen testbed snapshots with a reference implementation of the scoring rule,
    archived on Zenodo (Section~\ref{sec:artifact}).
\end{itemize}

\section{Background and Motivation}\label{sec:background}

\textbf{SBOMs are mandated, and the graph is part of the mandate.} Executive
Order 14028 made SBOMs a procurement expectation for software sold to the US
federal government~\cite{eo14028}, and NTIA's \emph{minimum elements} include
\emph{dependency relationships} among the baseline fields~\cite{ntia2021minimum};
CISA's follow-up framing keeps relationships in the baseline~\cite{cisa2024framing}.
The EU Cyber Resilience Act requires an SBOM covering at least the top-level
dependencies of a product but enumerates no relationship field~\cite{eu2024cra}. We measure
below how often the field is absent as we map it to each format; a document with a
single component has no relation to declare.

\textbf{Consumers read the graph, not just the list.} The scenarios that
motivate SBOM adoption are graph traversals. Vulnerability triage
pipelines~\cite{iot2026sbomtriage}
combine severity and exploitation signals (CVSS, EPSS~\cite{jacobs2021epss},
CISA KEV) whose disagreements are well documented~\cite{koscinski2025conflicting,
epsskev2024}, and then contextualize them with \emph{structure}: is the
vulnerable component reachable from the application, how central is it, what
is the blast radius~\cite{baird2026attackchains}. Impact and
propagation studies in Maven quantify how vulnerabilities travel along
dependency edges~\cite{ripple2025maven,lifecycle2025transitive,tracing2025cve,
propagation2025impact}; counting \emph{actually} vulnerable dependencies,
as opposed to inflated flat counts, requires knowing which dependencies are
real and how they connect~\cite{pashchenko2018vulnerable,pashchenko2022vuln4real}.
Supply-chain attack taxonomies likewise assume transitive visibility: the
attacks exploit exactly the paths that flat inventories do not
show~\cite{ohm2020backstabber,ladisa2023taxonomy,zimmermann2019smallworld}.
Every one of these consumers presupposes that the SBOM's edge set
approximates the truth. This paper measures how often that assumption has
any data behind it.

\section{Dataset and Method}\label{sec:method}

\textbf{Corpora.} We use three corpora, each for a different purpose. The
\emph{wild corpus} carries every population claim and answers RQ1--RQ4: Wild
SBOMs~\cite{soeiro2025wildsboms}, 78,612 unique SBOM files mined from Software
Heritage (1,782 forges), checked into repositories by their producers, with no
filtering on popularity or activity. One generation harness dominates: 36,551
files (46.5\%) carry an entry from the tooling of Fortress Information Security, almost always its \texttt{Github Extractor} 1.15.1, alongside up to six other tools, and an
independent second measurement traced them to a single automated harness that ran
several SBOM generators over public repositories in
2022~\cite{markovian2026sm003}. We report headline shares with and without it.
The \emph{image scans} isolate one archetype of degenerate graph: 11 SBOMs we
generated with Syft~\cite{anchore2024syft} (ten with 1.42.3, one with 1.42.4,
default configuration) from large public images (ML/AI, CI, data science,
distributed processing; 5,263--87,776 components). The selection is purposive, so the corpus
supports an archetype claim and no population estimate. The \emph{testbed} is 18
SBOMs (twelve public container images and applications, four applications built
to be vulnerable, two in-house applications) processed by a multi-signal
vulnerability-prioritization system, the only scoring system in this study; we call
its computation, detector and parser \emph{production} ones to set them apart from
the reference implementation in the deposit, not to imply that the system is
distributed. It shows the mechanism behind RQ5 and RQ6 and supports no population
claim.

\textbf{Measures.} For each SBOM we parse CycloneDX (JSON, XML) or SPDX (JSON, tag-value, YAML) and build the declared graph: nodes are components (\texttt{bom-ref} /
\texttt{SPDXID}), edges are declared dependency relations
(\texttt{dependencies[].dependsOn}; \texttt{DEPENDS\_ON}/\texttt{CONTAINS}
relationships). We record $V$, $E$, $E/V$, orphan ratio, connected components,
format, producing tool, the ecosystem mixture of its components (\texttt{purl} types), purl/version/scope coverage, and
the CycloneDX \texttt{compositions} completeness declaration, and we join each
SBOM to its sbomqs~\cite{interlynk2024sbomqs} score shipped with the corpus
metadata. The scanner streams the 5\,GB zstd-compressed archive in a single pass
and resumes idempotently after interruption.

The scanner and the statistical scripts were implemented with the assistance of
a generative-AI coding tool (see the AI declaration); we treat them as we would
any software instrument and validate accordingly. Two independent full scans run
three months apart, on different scanner revisions, produced bit-identical $V$
and orphan ratios for all 77,092 parseable SBOMs; all released code is open for inspection (Section~\ref{sec:artifact}).

\textbf{Taxonomy.} Let $G(S) = (V, E)$ be the directed graph declared by SBOM
$S$. The \emph{orphan ratio}
\begin{equation}
\rho(S) \;=\; \frac{\lvert \{\, v \in V : \deg_G(v) = 0 \,\} \rvert}{\lvert V \rvert}
\label{eq:orphan}
\end{equation}
uses the undirected degree, so an orphan neither depends on nor is depended upon
by any declared component. With $D$ the predicate ``$S$ declares a usable
dependency block'' (a non-empty \texttt{dependencies} array for CycloneDX; a
non-empty set of resolvable dependency relationships for SPDX), the regimes are
\begin{equation}
\mathrm{class}(S) =
\begin{cases}
\textsc{no\_edges} & \neg D,\\
\text{degenerate-with-edges} & D \,\wedge\, \rho(S) > \tau,\\
\text{connected} & D \,\wedge\, \rho(S) \le \tau,
\end{cases}
\label{eq:taxonomy}
\end{equation}
with $\tau = 0.5$. The regimes are separated by the block, not by the edge
count: 4,855 SBOMs, all CycloneDX, declare a block whose entries resolve to no
usable edge, sit at $\rho = 1$ and are classified degenerate (71.4\% of that
class). A block-less document is not treated as the $\rho = 1$ limit, because
omitting the field and declaring it empty are different acts. The population
therefore carries two measurands: 40,756 documents (52.9\%) lack a block, and
with the empty blocks 45,611 (59.2\%) realize no edge.

\textbf{Three measures of isolation.} We keep three apart. The taxonomy uses
\emph{full isolation} (undirected degree zero on the declared graph). A
degeneracy detector can use the weaker \emph{in-degree} rule, under which a
component is an orphan when nothing depends on it; on the same declared graphs
this leaves 63.9\% of parseable documents without an informative graph, against
61.7\% for the taxonomy. The detector of the testbed system applies the in-degree
rule to the \emph{parsed} graph, in which the parser links the document root to
every declared dependency entry that nothing else points to; it is defined only
for CycloneDX documents with a root (39,851 documents, 51.7\% of the parseable
population). The two in-degree measures disagree on 1,673 of these documents
(4.2\%), always in one direction (the declared graph calls degenerate what the
parsed graph calls informative), so the declared-graph share is an upper bound on
the parsed one. Section~\ref{sec:population} uses the first measure and
Section~\ref{sec:trap-population} the third.

\textbf{Generator attribution.} We attribute a document to the tool that produced
its content, not to the first entry of the tool chain, because two kinds of
entries are not producers: a generation harness (the Fortress entry, which we
treat as a flag unless it is the only entry) and a \texttt{Node.js module} entry
placed before another tool, which in these chains is a wrapping layer. One rule
was added after we saw the first result: the 1,579 documents whose chain holds only
the harness and a \texttt{Node.js module} entry are attributed to the harness.
The harness flag is a lower bound, since a harness that leaves no entry is
invisible to us.

\textbf{Threats.} (i) Wild SBOMs are repository-committed artifacts, not
build-fresh outputs; they reflect what producers publish, which is the
consumer-visible population, but may lag generator versions and underrepresent
consumer-side image scans. (ii) Attribution depends on self-reported metadata, and its rules were written by one person and not checked by a second; near-identical strings can denote different tools:
\texttt{Github Extractor} is Fortress Information Security's tool, whereas
GitHub's own dependency-graph exporter (\texttt{GitHub.com-Dependency-Graph})
appears in only 17 documents, 16 of them with edges. The harness makes the corpus
less independent than its size suggests, so every population share is also given
without it. (iii) SPDX \texttt{CONTAINS} relations count as edges, and in Syft's SPDX output many of them link packages to files rather than to dependencies; counting them can only move documents out of \textsc{no\_edges}, so with respect to these relations 52.9\% is a lower bound. (iv) Small SBOMs dominate the degenerate class by count (median
$V=3$, and 10,721 documents have at most one component); we therefore report
strata and make no claim about the software itself. (v) The variance decomposition is correlational and we make no causal claim. (vi) The KEV catalogue is incomplete (Section~\ref{sec:trap-population}), and the testbed
shows a mechanism, not a rate (Section~\ref{sec:trap}).

\section{Results I: The Population}\label{sec:population}

\subsection{RQ1: The Declared Graph}\label{sec:rq1}

Of 78,612 files, 77,092 (98.1\%) parse as CycloneDX (80\%) or SPDX (20\%). The
1,520 failures are dominated by files that are not SBOM documents at all: 76.1\%
are binary, and with unrecognized text, gzip, zip and bare hashes they make up
94.1\% of the residue. The exclusion therefore removes non-documents rather than a
class of graphs, though we cannot prove that the remaining 5.9\% (90 files, 0.11\%
of the corpus) is topology-neutral. Table~\ref{tab:taxonomy} shows the three-way
split. Of the parseable documents, 52.9\% omit the NTIA dependency-relationship
field entirely as we map it to the block of each format, in CycloneDX (53.7\%) and SPDX (49.5\%) alike. Counting dependencies expressed only by nesting components would lower the share to 52.8\% (85 documents). The generation harness accounts for at least 36,551
documents, 64.6\% of which lack the block; without them ($n = 40{,}541$) the three
shares are 42.3\%, 11.4\% and 46.3\%, and 53.7\% lack an informative graph.
Among documents with at least two components, 48.1\% lack the block and 55.5\%
lack an informative graph (39.4\% with the harness also set aside).

\begin{table}[t]
\caption{Declared-graph taxonomy over 77,092 parseable SBOMs.}
\label{tab:taxonomy}
\centering\footnotesize\setlength{\tabcolsep}{2.5pt}
\begin{tabular}{lrrrr}
\toprule
Class & $n$ & share & med.\ $V$ & med.\ $E/V$ \\
\midrule
\textsc{no\_edges} & 40,756 & 52.9\% & 16 & 0.000 \\
degenerate-with-edges & 6,796 & 8.8\% & 3 & 0.000 \\
connected & 29,540 & 38.3\% & 49 & 0.995 \\
\bottomrule
\end{tabular}
\end{table}

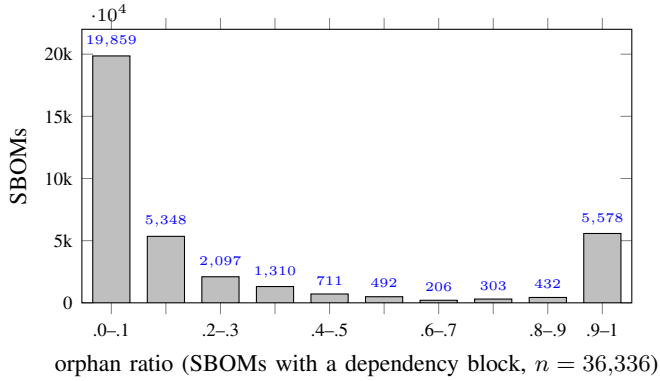
\begin{figure}[t]
\centering
\begin{tikzpicture}
\begin{axis}[
  ybar, bar width=14pt, width=\linewidth, height=5.2cm,
  ymin=0, ymax=22000, scaled y ticks=false,
  ylabel={SBOMs},
  xlabel={orphan ratio (SBOMs with a dependency block, $n=36{,}336$)},
  xtick={0.05,0.15,0.25,0.35,0.45,0.55,0.65,0.75,0.85,0.95},
  xticklabels={.0--.1,,.2--.3,,.4--.5,,.6--.7,,.8--.9,.9--1},
  ytick={0,5000,10000,15000,20000},
  yticklabels={0,5k,10k,15k,20k},
  tick label style={font=\scriptsize},
  label style={font=\small},
  nodes near coords, every node near coord/.append style={font=\tiny, text=black},
  enlarge x limits=0.06,
]
\addplot+[fill=black!25, draw=black] coordinates {
  (0.05,19859) (0.15,5348) (0.25,2097) (0.35,1310) (0.45,711)
  (0.55,492) (0.65,206) (0.75,303) (0.85,432) (0.95,5578)};
\end{axis}
\end{tikzpicture}
\caption{Orphan-ratio distribution over the block-declaring subpopulation:
a connected mode, a degenerate mode, and a near-empty valley (0.5--0.8 holds
2.8\% of the mass). Bins are left-closed, $[a, a{+}0.1)$, with the last bin
$[0.9, 1.0]$; the 215 SBOMs at exactly $\rho = 0.5$ fall in the 0.5--0.6 bar
yet classify as \emph{connected} under the strict $\rho > \tau$ rule, which is
why the bars from 0.5 upward sum to 7,011 against a degenerate-class size of
6,796. Of the 5,578 documents in the last bar, 4,855 declare a block that yields
no edge. \textsc{no\_edges} SBOMs (40,756) are excluded by
construction.}
\label{fig:histogram}
\end{figure}

\textbf{The degeneracy threshold.} The cut $\tau$ decides whether, for a document
as a whole, the absence of a path is the rule rather than the exception. We
fixed it on the testbed, whose 18 orphan ratios leave nothing between 0.33 and
0.93. Over the 36,336 block-declaring documents the distribution looks bimodal
(Fig.~\ref{fig:histogram}): 54.7\% below 0.1, 15.4\% at or above 0.9, and 2.8\% in
the whole 0.5--0.8 band. The upper mode, however, consists mostly of empty
blocks: 4,855 of its 5,578 documents declare a block that yields no edge, and
among the 31,481 documents with at least one edge only 2.3\% sit at or above 0.9.
The population therefore does not single out $\tau$, and the choice rests on the
reading above. Moving $\tau$ across $[0.4, 0.9]$ changes the share of documents
without an informative graph only from 62.4\% to 60.1\%, because missing and empty
blocks dominate it; the degenerate share among documents with edges, by contrast,
falls from 7.9\% to 2.3\%, and under the production detector of RQ4.1 from 56.8\%
to 53.2\%. The headline is insensitive to $\tau$; the size of the degenerate class
is not. Large documents are less affected: among those with $V \ge 50$, 1,370 of
the 1,582 degenerate ones realize at least one edge.

\textbf{Answer to RQ1.} Most SBOMs committed to public code give a consumer no
informative dependency graph: 52.9\% omit the dependency-relationship field,
although the NTIA minimum elements require it, another 8.8\% declare a graph in
which most components are orphans, and 38.3\% are connected. Set the harness
aside, and 42.3\% still lack a dependency block.

\subsection{RQ2: What Predicts the Regime}\label{sec:rq2}

The missing-block rate spans the whole scale across producing tools
(Table~\ref{tab:generators}). The harness's \texttt{Github Extractor} never emits
edges; Gemnasium (94.3\%), \texttt{cdxgen} (74.1\%) and Syft (58.1\%) usually omit
the block, while the lockfile-based \texttt{cyclonedx-gomod} (0.0\%) and
\texttt{cyclonedx-php-composer} (0.2\%) and Microsoft's \texttt{sbom-tool}
(4.3\%) almost always emit a graph. Two SBOMs of the same project, produced by
different mainstream tools, can therefore land in opposite regimes, and nothing in
a document tells the consumer which regime it is in. For a consumer, prediction is
enough: knowing which tool wrote the document narrows down sharply which regime it
is in, even though our design does not license a causal reading.

\begin{table}[t]
\caption{Regime shares by producing tool (the ten largest; 2,074 documents with
no tool entry omitted). \emph{Harness}: share of the tool's documents that carry
the Fortress harness entry. Microsoft \texttt{sbom-tool} declares edges only from
the root, which makes its documents connected under our rule whatever the depth of
the real graph. \texttt{cyclonedx-bom} (SPDX): SPDX conversions of CycloneDX
documents; all 964 JSON documents of this row list the converter
\texttt{CycloneToSpdx} 0.1.4 as a creator.}
\label{tab:generators}
\centering\footnotesize\setlength{\tabcolsep}{2.5pt}
\begin{tabular}{lrrrr}
\toprule
Producing tool & $n$ & harness & \textsc{no\_edges} & degen. \\
\midrule
\texttt{cdxgen} & 24,460 & 68.4\% & 74.1\% & 4.3\% \\
\texttt{cyclonedx-php-composer} & 7,125 & 94.4\% & 0.2\% & 16.4\% \\
\texttt{cyclonedx-gomod} & 6,521 & 86.9\% & 0.0\% & 11.6\% \\
\texttt{syft} & 4,997 & 0.1\% & 58.1\% & 5.4\% \\
\texttt{Github Extractor} (harness) & 4,630 & 100.0\% & 100.0\% & 0.0\% \\
Microsoft \texttt{sbom-tool} & 3,542 & 0.0\% & 4.3\% & 0.0\% \\
\texttt{cyclonedx-bom} (SPDX) & 2,891 & 0.0\% & 100.0\% & 0.0\% \\
Gemnasium & 2,705 & 0.0\% & 94.3\% & 0.0\% \\
\texttt{trivy} & 1,816 & 0.0\% & 42.6\% & 41.2\% \\
CycloneDX Maven plugin & 1,785 & 24.9\% & 3.0\% & 16.0\% \\
\bottomrule
\end{tabular}
\end{table}

Table~\ref{tab:eta} sizes the claim with $\eta^2$, the share of variance in a
binary outcome explained by group membership; it is a descriptive effect size,
not a test, since documents are not independent draws (half of the corpus enters
through ten bulk additions). The specification format explains essentially
nothing ($\eta^2 = 0.001$ for the missing block, 0.010 for $E = 0$). Grouping by
producing tool explains between 0.409 and 0.648, depending on how finely tools are
distinguished, each value more than ninety times its permutation baseline. Within
a tool, the writer for each format matters: the \texttt{trivy} family omits the
block in 90.3\% of its SPDX documents and in 4.4\% of its CycloneDX ones, the Syft
family in 13.5\% and 82.6\%. We take $\eta^2 = 0.588$ (format $\times$ family, 20
groups) as the reference value. The two outcomes are distinct measurands:
\texttt{trivy} in CycloneDX ($n = 1{,}057$) omits the block in only 4.4\% of its
documents yet leaves 75.1\% of them with no edge, declaring the required field and
filling it with entries that resolve to nothing.

\begin{table}[t]
\caption{Variance in edge emission explained by grouping ($\eta^2$, $n=77{,}092$).
\emph{Baseline}: largest $\eta^2$ over 200 random permutations of the group
labels, for either outcome.}
\label{tab:eta}
\centering\footnotesize\setlength{\tabcolsep}{2.5pt}
\begin{tabular}{lrrrr}
\toprule
Grouping & groups & baseline & no block & $E = 0$ \\
\midrule
format only (CycloneDX/SPDX) & 2 & 0.0001 & 0.001 & 0.010 \\
tool family & 15 & 0.0005 & 0.409 & 0.347 \\
format $\times$ family & 20 & 0.0005 & 0.588 & 0.471 \\
format $\times$ producer string & 440 & 0.0067 & 0.648 & 0.625 \\
\bottomrule
\end{tabular}
\end{table}

Because many tools serve a single ecosystem, tool and ecosystem are confounded. On
the 60,655 documents in which one ecosystem accounts for at least 90\% of the
components that carry a \texttt{purl}, the ecosystem alone accounts for $\eta^2 =
0.34$ of the missing-block variance and the tool family for 0.42, together 0.63;
for $E = 0$ the order narrowly reverses (ecosystem 0.38, tool 0.35, together
0.66). Without the harness, the tool accounts for 0.41 of the missing-block
variance and the ecosystem for 0.11. Tool releases explain much of this: outside
the harness, \texttt{cdxgen} omits the block in 18.1\% of its documents, not
74.1\%, and the harness and the \texttt{cdxgen} 3.x release it ran cannot be
separated. A tool thus emits edges for the ecosystems its builders cover, in the
releases that have them.

\textbf{Answer to RQ2.} The format does not predict the regime ($\eta^2 \le
0.010$), but the generating tool does: $\eta^2 = 0.41$ for tool family and 0.59
once each tool's format writer is distinguished, against permutation baselines
below 0.007. On documents that describe a single ecosystem, tool and ecosystem act
jointly (together 0.63). Outside the harness, the tool accounts for two to four
times as much as the ecosystem. This is a statement about prediction, not about
cause.

\subsection{RQ3: Whether a Consumer Can Tell}\label{sec:rq3}

Only 59 of 61,691 CycloneDX SBOMs (0.10\%) carry a \texttt{compositions} block,
and 54 of those declare a mixture that already contains \texttt{unknown}. Over the
whole corpus, the median sbomqs score of documents without a block (6.40) is
within 0.1 of that of connected ones (6.50), and the score separates the two with
an AUC of 0.60. That near-equality comes from the harness, within which the score
does not separate the regimes at all (AUC 0.49). Outside it the AUC is 0.79, but
part of that holds by construction: every sbomqs release current when the corpus
was scored (v0.1.9 to v0.2.3) averages 23 criteria, one of which awards full
points when the primary component declares a dependency. Recomputed without that
criterion, the AUC outside the harness is 0.72 against documents without a block
and 0.61 against degenerate documents, which do declare the block and so cannot
be told apart by checking for it. Quality frameworks answer ``how complete is this
document overall'', whereas a reachability consumer needs to know whether the
document carries a traversable graph.

\textbf{Answer to RQ3.} Not from completeness declarations, which almost no
producer emits, and only weakly from a quality score: 0.60 over the whole corpus,
0.49 within the harness, and, outside the harness, 0.72 against documents without
a block and 0.61 against degenerate ones once the score's own dependency criterion
is removed.

\section{Results II: What the Closed-World Reading Removes}\label{sec:trap-population}

If a consumer applies the veto to the documents of Section~\ref{sec:population},
what does it take away? We answer on the whole corpus, because the events of
interest are scarce. Of the 1,653 entries of the KEV catalogue of 2026-07-22, only
155 map to a package record in OSV through the CVE identifier and its aliases, a
route that does not reach distribution packages. We assign the regime with the
production detector (Section~\ref{sec:method}); in degenerate documents we count
the components that nothing depends on in the parsed graph, a subset of those
without a path from the root, so these counts are conservative, and in informative
documents the components without a directed path from the root.

\subsection{RQ4.1: The Degenerate Regime}\label{sec:rq61}

On the neutral measure of Section~\ref{sec:population}, a missing path carries no
information in 61.7\% of parseable documents. Restricted to what the production
detector can ingest, the share is 55.2\% (21,995 of 39,851 documents). In a
degenerate document the veto removes almost the whole inventory: the median
document would lose 96.4\% of its components (interquartile range 85.7--99.6\%).

We intersect the KEV-to-package table with the components the veto would suppress
and check every match against OSV at the component's exact version. This yields
29 distinct KEV-listed CVEs in 1,901 documents (2,584 instances): 8.6\% of
degenerate documents, against 0.9\% of informative ones. Matching by package name
alone gives 67 CVEs in 4,315 documents (7,543 instances), an upper bound on what
the version check can confirm; 189 instances could not be checked because their
version is missing or not a release number, or OSV does not resolve the package.
The most frequent name match, CVE-2025-54313 (2,186 instances), concerns a few
malicious releases of two npm packages and is confirmed in none. The table
contains no distribution packages, so the confirmed counts understate what a
complete inventory would show. The ground truth is a catalogue, and catalogues are
incomplete: a capture--recapture estimate~\cite{petersson2004capturerecapture}
that we computed from the APT campaign data of Di Tizio et
al.~\cite{ditizio2023updates}, which end in 2020, puts the sensitivity of the KEV
catalogue at $c \approx 0.895$, itself an upper bound. We therefore report counts
of catalogued instances, not of exploited vulnerabilities present in the corpus.

\textbf{Answer to RQ4.1.} Where the graph is uninformative, reading a missing path
as proof of unreachability would discard most of the inventory, and catalogued
exploited vulnerabilities with it: 29 distinct version-confirmed KEV-listed CVEs in
1,901 documents (by package name alone, 67 CVEs in 4,315 documents). Components
without a package identifier and distribution packages are not covered, so these
are lower bounds.

\subsection{RQ4.2: The Informative Regime}\label{sec:trap-f4}

If silence must map to \emph{unknown} where extraction failed, the mirror question
is what the veto does where extraction appears to have succeeded. Checked at the
exact version in the same way, in the informative regime it covers 14 distinct
KEV-listed CVEs in 164 documents (203 instances; by package name alone, 27 CVEs in
545 documents; 79 instances could not be checked): about one-twelfth as many
documents but half as many distinct CVEs. The affected CVEs also differ in kind:
HTTP/2 Rapid Reset, Spark, Spring4Shell and two Tomcat vulnerabilities lead the
informative regime. The two counts are not symmetric in construction, since in
degenerate documents we count only components that nothing depends on, so the
difference between the regimes is if anything understated.

This does not show that the veto is wrong in the informative regime. The
measurement establishes the \emph{extent} of the veto there, not its
\emph{correctness}: where a graph really is informative, absence of a path is
intended to be evidence, and if those components are never loaded, the veto is
right. Separating the two requires code-level reachability, which a corpus of SBOM
documents cannot supply. The analysis plan of this measurement was written before
it was run, and version~1 of this preprint attests its question and parameters.

\textbf{Answer to RQ4.2.} Where the graph is informative, the veto touches far
fewer documents but not proportionally fewer vulnerabilities: 14 version-confirmed
KEV-listed CVEs in 164 documents (0.9\% of informative documents), against 29 in
the degenerate regime. This tells us how far the veto extends there, not whether it
is right.

\section{Image Scans: The Large-Degenerate Archetype}\label{sec:containers}

Across the 11 image scans (nvidia/cuda, pytorch, tensorflow,
jupyter/all-spark-notebook and GitLab CE, among others; $V$ from 5,263 to
87,776), every SBOM lands deep in the degenerate regime (Table~\ref{tab:images}):
91.4--98.0\% of components are orphans, at $E/V$ between 0.026 and 0.229. Most of
that orphan mass is made of components of type \texttt{file}, which carry no
\texttt{purl} and appear in no dependency entry. Counted over packages alone, the
components that carry a \texttt{purl}, the orphan share is 1.1--49.1\% (35.7\%
pooled). The images thus reach the degenerate regime by a different route than the
population, where 99.0\% of the 10,778,582 components that nothing depends on
carry a \texttt{purl}: in the images the scanner lists files without relations,
whereas in the population tools omit the relations between packages. The detector
does not distinguish the two routes, and here the file entries decide the regime.
Counted over packages alone, none of the 11 images would be degenerate, and a
closed-world consumer would set aside at least the isolated packages, up to 49.1\%
of them in \texttt{devcontainers/universal}.

\begin{table}[t]
\caption{The 11-image corpus (Syft 1.42.3/1.42.4 scans of public images).
\emph{orph.}: orphan share over all components; \emph{orph.\ pkg}: over the
components that carry a \texttt{purl}.}
\label{tab:images}
\centering\footnotesize\setlength{\tabcolsep}{2.5pt}
\begin{tabular}{lrrrrr}
\toprule
Image & $V$ & $E$ & $E/V$ & orph. & orph.\ pkg \\
\midrule
gitlab/gitlab-ce:17.0 & 87,776 & 2,306 & 0.026 & 97.6\% & 28.7\% \\
devcontainers/universal:2 & 67,623 & 5,629 & 0.083 & 94.7\% & 49.1\% \\
jupyter/all-spark-notebook & 41,628 & 2,206 & 0.053 & 97.9\% & 45.0\% \\
rocker/tidyverse & 19,093 & 1,182 & 0.062 & 97.9\% & 35.4\% \\
catthehacker/ubuntu:act & 18,371 & 2,732 & 0.149 & 91.5\% & 20.5\% \\
tensorflow:2.16.1-gpu & 13,392 & 993 & 0.074 & 97.5\% & 7.3\% \\
pytorch:2.3.0-cuda12.1 & 11,214 & 781 & 0.070 & 97.1\% & 5.8\% \\
nvidia/cuda:12.4.0-devel & 11,078 & 618 & 0.056 & 98.0\% & 2.2\% \\
continuumio/anaconda3 & 8,744 & 2,002 & 0.229 & 93.9\% & 8.7\% \\
apache/spark:3.5.1 & 6,020 & 649 & 0.108 & 94.3\% & 40.1\% \\
gitlab/gitlab-runner & 5,263 & 686 & 0.130 & 91.4\% & 1.1\% \\
\bottomrule
\end{tabular}
\end{table}

The \texttt{apache/spark:3.5.1} document shows the split: 5,450 of its 6,020
components are of type \texttt{file} (JAR, shared-object and Python files whose
owning package was not resolved), and none of them appears in a dependency entry.
We say \emph{image-scan} SBOMs, not container SBOMs in the wild. Of the 2,718
repository-committed documents our heuristic flags as container-related, only 9
(0.3\%) are degenerate-with-edges, because they are mostly derived from
Dockerfiles or source rather than from built images; the corpus therefore supports
an archetype claim and no estimate of how common the archetype is among the
documents consumers read. Prior container-security studies counted vulnerabilities
in images at scale~\cite{shu2017dockerhub,zerouali2019relation}; to our knowledge
none examined the structure of the graph that image SBOMs declare.

\section{The Testbed: Mechanism and Illustration}\label{sec:trap}

\textbf{The unsound inference.} A vulnerability-prioritization pipeline can multiply a reachability factor into
its score, as the system we study does (how widely the practice is deployed, a
public corpus cannot reveal), and \emph{unreachable} then acts as a veto
($\times 0$). On a declared graph this encodes the closed-world
assumption~\cite{reiter1978closed}: from ``no declared path'' infer ``no path''.
SBOM extraction is \emph{soundy}~\cite{livshits2015soundiness} with a specific
asymmetry: declared edges are trustworthy (extractors transcribe manifests; they
do not invent dependencies, an assumption we state rather than measure), but absent
edges are not evidence (Sections~\ref{sec:population}--\ref{sec:containers}).
Silence must therefore map to \emph{unknown}, not \emph{false}; the veto remains
legitimate only for \emph{evidence-based} unreachability, an excluded scope or
isolation in a graph the detector classifies as informative.

\textbf{The testbed and its ground truth.} Each advisory matched to a component is
scored from CVSS, EPSS, component importance and the reachability factor, and
raised by KEV membership, threat-actor match and exploitation status; the score is
mapped to an SSVC-style decision, and a match is \emph{actionable} when the
decision is \emph{act} or \emph{attend}. Mission criticality comes from
organization profiles written for the testbed, so it is an assumption, not a
measurement, and the profiles are not deposited. Ground truth is the CISA KEV
catalogue at the snapshot date (1,617 entries for the offline snapshot of
2026-06-10, 1,631 for the live run of 2026-07-03). We score with the KEV flag
removed; EPSS, which itself encodes catalogue information, is retained. Because
matching by package name can attach an advisory to a version it does not affect,
we re-checked every match against OSV with the exact package version: of the 4,273
matches of 2026-07-03, 2,813 are confirmed, 804 refuted and 656 unresolved
(offline snapshot: 2,450, 325 and 595 of 3,370). Testbed figures use confirmed
matches unless stated otherwise. Each snapshot has 14 confirmed KEV-listed matches, 11 of them in the deliberately vulnerable applications. The analysis plan was written after the baseline measurement, and the runs depart from it in the detector metric, in how alert volume is measured and in the reference arm.

\textbf{What the veto did.} On the offline snapshot, scored with the KEV flag
removed, the veto suppressed 5 of the 14 KEV-listed matches whose package version is
confirmed. All five are in \texttt{log4shell-app} and all sit on
components without a path from the root, among them Log4Shell (CVE-2021-44228) on
\texttt{log4j-core} and Spring4Shell (CVE-2022-22965) on \texttt{spring-beans}. On all
matches, including unconfirmed ones, the veto accounts for 7 of the 8 missed
KEV-listed matches; the two beyond \texttt{log4shell-app}, in the Grafana image,
rest on matches whose version could not be confirmed. Mapping \emph{unreachable} to
\emph{unknown} in documents the detector classifies as degenerate restores all five.

\textbf{Why this is an illustration.} The direction of this change is fixed by
construction: raising a factor from 0 to 0.5 cannot lower a score, and the decision
is a non-decreasing step function of the score, so the actionable set after the
change contains the one before. Only the size of the change is an observation, and
it rests on a single document. \texttt{log4shell-app} is degenerate only through its
file entries: over packages alone its in-degree orphan ratio is exactly 0.5, which
the detector's strict threshold classifies as informative. The \emph{unknown}
factor was chosen on the same vetoed instances whose restoration we report, 11 of
the 14 instances come from deliberately vulnerable applications, and EPSS retention
leaves the suppression unaffected but may help the restored matches cross the
decision threshold. We run no significance test and claim the mechanism, not a rate
or a size.

\textbf{Answer to RQ5 and RQ6.} In the 18-SBOM testbed, scored without the KEV flag,
the closed-world reading suppressed 5 of the 14 KEV-listed instances with a
confirmed package version, all in one deliberately vulnerable application, and
treating unreachable components as unknown in documents the detector flags as
degenerate brings all five back. The result is fragile and illustrates the
mechanism; it does not measure its effect.

\textbf{The unused escape hatch.} CycloneDX \texttt{compositions} lets a producer
state that the declared graph is incomplete~\cite{cyclonedx2024spec}; consumers
could route semantics on it. We find 0.10\% usage (59 of 61,691 CycloneDX
documents, most of them declaring a mixture that includes \texttt{unknown}).
Producers rarely declare incompleteness, and the specification does not require them to, which leaves consumers without a machine-readable signal; until that changes, the open-world reading is the only safe default.

\section{Implications}\label{sec:implications}

\textbf{For producers and toolmakers.} The regime an SBOM lands in is predicted by
the generating tool and its writer for each format (Tables~\ref{tab:generators} and~\ref{tab:eta}); where edges are recoverable, tools
that read resolved lockfiles already recover them. Where edges genuinely cannot be recovered,
emitting a one-line \texttt{compositions} declaration would convert silent
incompleteness into machine-readable uncertainty; today that
signal exists in 0.10\% of CycloneDX documents. An empty block is the worse option,
since it satisfies a compliance check while conveying nothing; 4,855 documents carry
one.

\textbf{For specifications and quality gates.} The NTIA dependency-relationship
field is absent from 52.9\% of parseable documents as we map it (42.3\% without the
generation harness). Field-oriented quality scores separate the regimes only weakly
once their own dependency criterion is removed (AUC 0.72 and 0.61 outside the
harness). Quality frameworks should score the \emph{graph} itself (presence of an
edge block, orphan ratio, a completeness declaration) as a first-class dimension
alongside field population.

\textbf{For consumers.} The single inference our results forbid is \emph{no declared
path} $\Rightarrow$ \emph{unreachable}. We propose to treat silence as uninformative
and to keep the veto for evidence: a degeneracy detector on the in-degree orphan
ratio decides whether the graph is informative, components without a path in a
degenerate graph become \emph{unknown}, and the veto is kept for evidence-based
unreachability. The principle treats declared edges as evidence, an assumption
about extractors, and it needs no better data. Our testbed shows the mechanism at
work; it does not measure how much the principle recovers in general.

\section{Related Work}\label{sec:related}

\textbf{SBOM quality at scale.} Adherence-gap studies audit fields and standard
compliance: 55K SBOMs, inter-tool package-detection consistency as low as
7.8--12.8\%~\cite{wang2026adherence}; tool comparisons show divergent container
results~\cite{bufalino2025sbomproof}, generator-dependent vulnerability assessment
in Python~\cite{benedetti2024python}, and NTIA minimums unmet in
practice~\cite{halbritter2024ntia}; standard-level and tool-landscape analyses
contrast the SPDX and CycloneDX ecosystems~\cite{bangash2025spdxcdx,sbomlandscape2024},
and SBOM distribution itself lacks integrity protection~\cite{sbomintegrity2024};
flat SBOMs are known to inflate false positives downstream~\cite{zhou2025realitycheck}.
Corpus papers mine SBOMs at scale (Wild SBOMs~\cite{soeiro2025wildsboms}, Maven
Central SBOMs~\cite{gamage2025mavensboms}, SPDX documents on
GitHub~\cite{sbom100k2025}). One study measures the declared graph itself: Nocera
et al.\ built dependency graphs from 119 SBOMs of 84 projects and found a path from
the primary component to any other node in only 20 of 89 CycloneDX documents,
although the dependency-relationship field was present in
78~\cite{nocera2025adoption}. That is the phenomenon we measure at population
scale; their corpus is the set of projects that depend on a generator, not a
population of published documents, and we add the structural taxonomy and quantify
the absence of the NTIA dependency-relationship field population-wide.

\textbf{On-demand exports.} A large-scale study of GitHub-generated SBOMs reports
that ``100\% of SBOMs provide dependency relationships for all listed
components''~\cite{bhuiyan2026beyondcompliance}, which appears to contradict our
finding that 52.9\% of SBOMs in the wild lack the block. The two results describe
different populations. Bhuiyan et al.\ generate 10,000 SBOMs on demand through
GitHub's dependency-graph endpoint for the most-starred repositories in ten
languages, restricted to active projects with at least five contributors; we
analyze documents that producers committed, with no selection on popularity,
activity or tooling. Their object is what one service \emph{can produce}; ours is
what was \emph{shipped}. Our corpus reproduces their result on their generator: of
the 17 documents created by \texttt{GitHub.com-Dependency-Graph}, 16 carry a block,
and all 16 are trees ($E = V-1$), which fixes the shape but not the depth of a graph, so we make no claim about depth in either population. The similarly named \texttt{Github Extractor}, the Fortress
harness tool, emits a block in none of its 4,630 documents, so two tools whose
names differ by one capital letter sit at opposite extremes of our taxonomy.

\textbf{Files-vs-metadata gap.} Software Dark Matter quantifies security-relevant
files invisible to SBOM metadata and filters triage by file
reachability~\cite{reddypalle2026sdm}. The two axes are orthogonal (undeclared
existence vs.\ declared-but-edgeless); both undermine closed-world consumption.

\textbf{Registry dependency networks.} Ecosystem-level topology is well studied:
npm/PyPI/CRAN evolution~\cite{decan2019empirical,kikas2017topology}, npm's
security-relevant small-world structure~\cite{zimmermann2019smallworld}, Maven
Central's 1.3M-node network~\cite{ogenrwot2025maven}, and vulnerability
propagation along registry edges~\cite{ripple2025maven,propagation2025impact}.
Those graphs are \emph{resolver-complete} reconstructions from registries; the
graph a consumer actually receives inside an SBOM artifact has, to our knowledge,
been characterized once before and on 119 documents~\cite{nocera2025adoption},
never at population scale. Outdatedness studies of Docker
images~\cite{zerouali2019outdatedjs} suggest the consumer-side artifacts are the
ones that lag the most.

\textbf{Code-level reachability.} Call-graph--based SCA (e.g., Eclipse
Steady~\cite{ponta2018beyond}) prunes false positives by proving vulnerable code
unreachable, and comparative studies document large inter-tool disagreement in what
SCA tools report~\cite{imtiaz2021comparative}. The veto sits a layer below: in
degenerate declared graphs, the component-level graph needed to even seed such
analysis is missing, and hidden code-level
dependencies~\cite{hidden2026dependencies} further widen the gap. Unsoundness of
``no path found'' verdicts mirrors known call-graph
incompleteness~\cite{reif2019missing}.

Our statements about gaps in prior work rest on a search run in July 2026 and
re-checked before the journal submission, across arXiv
(\texttt{cs.SE}, \texttt{cs.CR}), the venues where SBOM measurement work has
appeared (MSR, EMSE, ARES, ESEM, ICSME, SANER), the CycloneDX and SPDX
specification sites, the NTIA minimum-elements documents and grey literature on
reachability filtering. Items on which a claim turns were read at source rather than
from abstracts. It was not a systematic review in the PRISMA sense. One study
appeared between the search and the re-check, the analysis of on-demand generated
SBOMs discussed above, and we expect others.

\section{Artifact Availability}\label{sec:artifact}

We release the derived data in full, in two archived records, together with what a
reader needs to recompute every number in this paper. The scanner, the analysis
scripts behind every figure, the script that assembles the data record and a
reference implementation of the testbed's scoring rule are archived at
doi:10.5281/zenodo.23854692~\cite{zieba2026topologycode} under the MIT licence. The
per-document measures for all 78,612 files with all three orphan measures, the
distinct package identifiers, the catalogue-to-package table, the generator
taxonomy without which the variance decomposition cannot be recomputed, the
11-image corpus, the frozen testbed snapshots of 2026-06-10 and 2026-07-03 with
their expected outputs, and a manifest with SHA-256 for every item are archived at
doi:10.5281/zenodo.23854730~\cite{zieba2026topologydata} under CC-BY-4.0. Both are
version~1.1.0, which accompanies this version of the preprint. Version~3 linked the
first version of these records (doi:10.5281/zenodo.21901870 and
doi:10.5281/zenodo.21886189); versions~1 and~2 linked none. The primary corpus is
not redistributed and remains governed by the Wild SBOMs dataset
licence~\cite{soeiro2025wildsboms}. The testbed's upstream matching ran against live
services (NVD, OSV, OpenCTI, MISP) and cannot be re-executed to the same result;
from the frozen snapshots onward the reference implementation reproduces every
testbed figure. The organizational profiles behind mission criticality are not
deposited.

\section{Conclusion}

Between two fifths and a half of the SBOMs published in the wild declare no
dependency block at all, depending on whether one dominant generation harness is
counted; a further tenth declare a graph that is mostly isolated vertices. Which of
the two a consumer receives is predicted not by the format but by the tool that
wrote the document, jointly with the ecosystem where a document describes one, and
nothing in the document announces the difference. Consumers who read these graphs
closed-world turn missing data into confident negative verdicts: over the corpus,
the same inference would suppress components carrying 29 version-confirmed
known-exploited vulnerabilities across 1,901 documents, and in our own pipeline it
suppressed Log4Shell. We also measured the mirror case, where the veto is meant to be
evidence, and report its extent without claiming it is wrong. The repair is simple:
detect when the graph cannot support the inference, treat silence there as
\emph{unknown}, and keep the veto for evidence. Until producers declare completeness
and consumers respect it, ``no edges'' must mean ``no verdict.''

\section*{Acknowledgment}
We thank Markovian Protocol for an independent second measurement of this
work~\cite{markovian2026sm003}, which identified the vendor of the
\texttt{Github Extractor} entry and the generation harness behind nearly half of
the corpus.

\section*{Declaration of Generative AI and AI-Assisted Technologies}
During the preparation of this work the author used Claude (Anthropic), via the
Claude Code CLI (2026), for three purposes: assisting with drafting and language
of the manuscript, supporting the literature search, and helping implement the
analysis tooling described in Section~\ref{sec:method} (the streaming scanner
and the statistical scripts). Literature identified with AI assistance was
verified against primary sources before citation. All experiments, data, and
findings are the author's; the author reviewed and edited all content and takes
full responsibility for the content of the published article. The generative-AI
tool is not, and cannot be, an author. A complete record of the assisted
workflow is preserved in the project's version-control history.

\bibliographystyle{IEEEtran}
\bibliography{refs}

\end{document}